\documentclass[12pt,preprint]{aastex}

\usepackage[normalem]{ulem}
\usepackage{multicol, multirow, units, xspace}

\newcommand{\mir}{mid-IR\xspace}
\newcommand{\mirc}{Mid-IR\xspace}
\newcommand{\ngal}{348\xspace}
\newcommand{\avg}[1]{\left< #1 \right>} % for average

\shorttitle{Optical CMD of Compact Groups}
\shortauthors{Butterfield et al.}

\begin{document}

\title{The Optical Green Valley vs \mirc Canyon in Compact Groups}

\author{Lisa May Walker, Natalie Butterfield, Kelsey Johnson, and Catherine Zucker}
\affil{Astronomy Department, University of Virginia,
    Charlottesville, VA 22904}

\author{Sarah Gallagher}
\affil{Department of Physics and Astronomy, University of Western Ontario,
    London, ON N6A 3K7, Canada}

\author{Iraklis Konstantopoulos}
\affil{Australian Astronomical Observatory, PO Box 915,
    North Ryde NSW 1670, Australia}

\author{Ann Zabludoff}
\affil{Steward Observatory, University of Arizona,
    Tucson, AZ 85721}

\author{Ann E. Hornschemeier and Panayiotis Tzanavaris}
\affil{Laboratory for X-Ray Astrophysics, NASA Goddard Space Flight Center,
    Greenbelt, MD 20771}

\and

\author{Jane C. Charlton}
\affil{Department of Astronomy and Astrophysics, Pennsylvania State University,
    University Park, PA 16802}

\begin{abstract}
Compact groups of galaxies provide conditions similar to those experienced by galaxies in the earlier universe. Recent work on compact groups has led to the discovery of a dearth of mid-infrared transition galaxies (MIRTGs) in IRAC (3.6--8.0 $\mu m$) color space \citep{johnson07,walker12} as well as at intermediate specific star formation rates \citep{tzanavaris10}. However, we find that in compact groups these mid-infrared (\mir) transition galaxies in the mid-infrared dearth have already transitioned to the optical ([g-r]) red sequence. We investigate the optical color-magnitude diagram (CMD) of 99 compact groups containing \ngal galaxies and compare the optical CMD with \mir color space for compact group galaxies. Utilizing redshifts available from SDSS, we identified new galaxy members for 6 groups. By combining optical and \mir data, we obtain information on both the dust and the stellar populations in compact group galaxies. We also compare with more isolated galaxies and galaxies in the Coma cluster, which reveals that, similar to clusters, compact groups are dominated by optically red galaxies. While we find that compact group transition galaxies lie on the optical red sequence, LVL+SINGS \mir transition galaxies span the range of optical colors. The dearth of \mir transition galaxies in compact groups may be due to a lack of moderately star forming low mass galaxies; the relative lack of these galaxies could be due to their relatively small gravitational potential wells. This makes them more susceptible to this dynamic environment, thus causing them to more easily lose gas or be accreted by larger members.
\end{abstract}

\keywords{galaxies: evolution --- galaxies: interactions --- galaxies: clusters}

%%%%%%%%%%%%%%%%%%%%%%%%%%%%%%%%%%%%%%%%%%%%%%%%%%%
%	Introduction
%%%%%%%%%%%%%%%%%%%%%%%%%%%%%%%%%%%%%%%%%%%%%%%%%%%
\section{INTRODUCTION}
Given that sensitive and detailed observations of the early era (z $\sim$ 4) of galaxy assembly are currently unattainable, studying local analogs is a key path toward making progress. While galaxies have evolved significantly from those earlier times, compact groups at the present epoch nevertheless provide a unique opportunity to study environments with similarities to the earlier universe \citep{baron87}, due to their high number densities, low velocity dispersions and frequent interactions \citep{hickson92}.

Studies of compact groups have revealed that in this environment member galaxies are subject to intense and frequent gravitational torques that affect them in a variety of ways. For instance, as a class, compact group galaxies are known to be deficient in \ion{H}{1} relative to field galaxies \citep{verdes01,borthakur10}. In addition, fundamental characteristics of compact group galaxies, such as the overall impact of this environment on star formation, are still a puzzle. Different studies of star formation in compact groups yield disparate results; for example, \citet{delarosa07} find that stellar populations in compact group ellipticals are older than in field ellipticals, while studies such as \citet{gallagher10} report intense current star formation in Hickson Compact Group 31 illustrating the wide range of star formation seen in compact groups.

Characterizing the star formation activity of compact group galaxies is crucial to understanding how this environment affects galaxy evolution. \citet{johnson07} studied the mid-infrared IRAC (3.6--8.0 $\mu m$) colors of galaxies in 12 compact groups to understand their star formation properties. Their results were surprising: they found a ``gap'' in mid-infrared (\mir) color space between galaxies with colors of quiescent stellar populations and galaxies with colors indicative of strong star-forming activity. Further work by \citet{tzanavaris10} studied galaxies from 11 of the 12 compact groups and found a gap in their specific star formation rates (star formation rate per unit stellar mass).

\citet{walker12} extended the \citet{johnson07} study to a larger sample of 49 compact groups containing 174 galaxies. They found that the underdensity of \mir transition galaxies persists in a smaller region of \mir color space that they call the ``canyon.'' This dearth of galaxies is not present in comparison samples of isolated galaxies, the center of Coma, or interacting pairs. However, the Coma infall region shows a similar distribution in IRAC color space to compact group galaxies. This is interpreted as a similarity between the environment in compact groups and the Coma infall region, both having high densities and reservoirs of unprocessed gas.

The \mir colors of galaxies only reveal their current specific star formation activity as manifest in stellar light reprocessed by dust. In order to more rigorously investigate their recent star formation history, we must also consider their optical colors -- whether they fall in the red sequence, blue cloud, or green valley in the optical color-magnitude diagram (CMD). CMD galaxy distributions are known to be strongly dependent on galaxy environment. CMDs of field galaxies tend to have both a strong blue cloud of actively star-forming galaxies and a red sequence of ``red and dead'' galaxies, with an under density of galaxies falling in the green valley. This shape differs strongly from CMDs of cluster galaxies, which are dominated by the red sequence, with few galaxies in the blue cloud or green valley. Placing the compact group CMD within the context of CMDs from other environments can yield insight into galaxy evolution in these important environments.

One of the more interesting regions in a CMD is the green valley between active and quiescent galaxies \citep{wyder07,martin07}, thought to be a transition region of galaxies in which star formation has recently ceased \citep{thilker10}. As the young, blue, massive stars become more scarce, galaxies cross the green valley and enter the red sequence \citep{wyder07}. The underdensity of galaxies in the green valley is thought to be due to the short crossing time between the blue cloud and red sequence, which should be on the order of a B star's lifetime \citep{thilker10}. Comparison of the optical and \mir colors of compact group galaxies will reveal whether the \mir transition galaxies fall in the optical green valley. This would then indicate an intrinsic relationship between the ionizing UV photons from young OB stars and the heating of dust seen in the \mir. In particular, comparing the distribution of galaxies in the optical CMD and \mir colorspace can reveal if and how the transition of stellar populations is related to the transition of the interstellar medium.

To form a more comprehensive picture of star formation in compact groups, we have embarked on a study of the optical properties of 99 compact groups drawn from two catalogs, as discussed in section \ref{data}. In this paper, we present the optical CMD of this sample along with three comparison samples, and compare the optical and \mir colors to understand the evolution of star formation in compact groups. In particular, we wish to address the question of whether the \mir transition galaxies correspond to the optical green valley.

%%%%%%%%%%%%%%%%%%%%%%%%%%%%%%%%%%%%%%%%%%%%%%%%%%%
%	2. Data
%%%%%%%%%%%%%%%%%%%%%%%%%%%%%%%%%%%%%%%%%%%%%%%%%%%
\section{DATA}
%%Sample%%%%%%%%%%%%%%%%%%%%%%%%%%%%%%%%%%%%%%%%%%%
\subsection{The Sample}
%% Compact Groups %%
\subsubsection{Compact Groups}\label{data}
For this study we compiled a sample of groups from the Hickson Compact Group catalog \citep[HCG;][]{hickson82} and Redshift Survey Compact Group catalog \citep[RSCG;][]{barton96}. Since these compact groups are from two different surveys it is important to consider their different selection criteria and the effect that may have on our analysis.

HCGs were visually chosen based on three specific photometric criteria \citep{hickson82}. They had to have a population of four or more members within three magnitudes of each other (though redshift information later revealed that a number of groups actually only had three members), be sufficiently isolated, and be compact (based on a surface brightness criterion).

RSCGs were selected from a redshift survey, using a friends-of-friends algorithm \citep{barton96}. Galaxies were identified as neighbors based on projected separation ($\Delta D \le 50 \;\mathrm{kpc}$) and line-of-sight velocity difference ($\Delta V \le 1000 \;\mathrm{km\;s^{-1}}$). Linked sets of neighbors became groups. The aim was to identify groups similar to the HCGs; as a result some RSCGs are also HCGs. One significant change from the criteria used in \citet{hickson82} is the lack of an isolation requirement. Due to the absence of this criterion there arises the possibility of RSCGs being embedded in larger structures. The most extreme examples of this are RSCGs 67 and 68, embedded in the Coma cluster, and RSCG 21, embedded in the Perseus cluster; we have excluded these groups from our analysis. Other than the surrounding environment, RSCGs seem to be quite similar to HCGs, and their \mir colorspace distributions are consistent.

To be consistent with the sample from \citet{walker12}, we included all HCGs and RSCGs (excluding RSCGs 21, 67, and 68) at $z< 0.035$ (to ensure the \mir polycyclic aromatic hydrocarbon features remain in their rest-frame bands) that were available through the SDSS DR8 archive \citep{aihara11}. This selection process yielded 28 HCGs, 58 RSCGs, and 13 groups that are in both catalogs. Groups that appear in both the HCG and RSCG catalogs are considered only once in our analysis, and will be identified by their HCG designation. In addition, utilizing redshifts available from the SDSS archive, we identified new galaxy members for 6 groups. To be considered part of the group, we required that new members be located between original group members in both projected position and redshift.

%% Comparison Samples %%
\subsubsection{Comparison Samples}
In order to study the effect of galaxy environment, we compare the compact group sample with galaxies from the Coma cluster \citep{mahajan10} as well as two samples of field galaxies: the first from the low-z survey of the NYU value-added galaxy catalog \citep[VAGC;][]{blanton05,blanton05b,adelman08,padmanabhan08}; the second is LVL+SINGS \citep{dale07, dale09}, this sample provides the opportunity to compare optical and \mir data \citep[to be consistent with][we only include galaxies above $\log{\left(L_{4.5} \left[\unit{erg\;s^{-1}\;Hz^{-1}}\right]\right)} = 27.5$]{walker12}. To maintain consistency with our compact group sample, we only included galaxies at $z< 0.035$. Comparing the CMDs of these different samples will help us understand the effect of the compact group environment on the interstellar medium and star formation of galaxies.

%%Photometry%%%%%%%%%%%%%%%%%%%%%%%%%%%%%%%%%%%%%%%
\subsection{Completeness}
Because the samples used here do not have rigorous completeness limits, it is important to investigate what effect incompleteness in one or more of the samples might have on the results and interpretation. 
The HCG catalog is complete down to $m_G < 13.0$ \citep{hickson82}, while the RSCG catalog is magnitude limited to $m_{B_0} \le 15.5$ \citep{barton96}. The faintest systems in the LVL sample approach $m_B = 19$, but the majority of the sample is at $m_B < 15$ \citep{dale09}, while the SINGS galaxies span $-23.5 < M_R < -12.5$ \citep{dale07}. Both the VAGC and the Coma samples are constrained by the SDSS spectroscopy limit of $r < 17.77$ \citep{blanton05b,mahajan10}; we note that the VAGC sample is over 95\% complete over the relevant magnitude range except at very low surface brightness. For simplicity, we compare the $M_r$ distributions in Figure \ref{lumhist}. Despite not having rigorous magnitude limits, it is clear that the LVL+SINGS and compact group samples are well matched. For Coma, there is a more noticeable difference, as the compact group sample includes more low-luminosity galaxies. The difference with VAGC is pronounced, in particular a straight comparison of galaxy populations between this catalog and compact groups at $M_r < -19.5$ should be done with care.

%%Photometry%%%%%%%%%%%%%%%%%%%%%%%%%%%%%%%%%%%%%%%
\subsection{Photometry}
In addition to performing our own photometry, we also conducted a consistency check against the VAGC. As it was obtained as part of a survey, the Sloan photometry was performed in a fundamentally different manner. Rather than customizing the aperture shape for each individual galaxy, the Petrosian fluxes were measured using a circular aperture of 2 Petrosian radii. However, this does not always account for all the light from a galaxy; for instance, only 80\% of the flux is measured for a de Vaucouleurs profile \citep{blanton01}. For irregular morphologies, such as those found in compact groups, this problem is likely to be exacerbated. 

Photometry was performed with SURPHOT \citep{reines08}, which determines apertures based on contour levels in a reference image (we used a sum of the {\it gri} images as these were the filters of interest to our study), then applies the aperture to each image of interest. We converted SDSS magnitudes to AB magnitudes, then applied K-corrections to z = 0 using {\tt kcorrect v4\_2} \citep{blanton07b}. Reddening corrections utilized the Galactic dust maps of \citet{schlegel98}. Any obvious stars in the apertures were removed using SURPHOT; in most cases these had a negligible effect on the photometry. The low-z VAGC contains photometry for 21 of the galaxies in our compact group sample; a comparison reveals that the two methods are consistent {\bf($\mathrm{\avg{[g-r]}} = 0.026$, $\sigma_{[\mathrm{g-r}]} = 0.025$, $\mathrm{\avg{M_i}} = 0.16$, $\sigma_{\mathrm{M_i}} = 0.13$)}, as shown in Figure \ref{vssloan}. Thus, for the 21 compact group galaxies in the VAGC, we could use the SDSS photometry in this work. However, to maintain consistency within our sample, we use our custom photometry for all galaxies.

%%%%%%%%%%%%%%%%%%%%%%%%%%%%%%%%%%%%%%%%%%%%%%%%%%%
%	3. Results
%%%%%%%%%%%%%%%%%%%%%%%%%%%%%%%%%%%%%%%%%%%%%%%%%%%
\section{RESULTS}
The main results of this paper are shown in Figure \ref{cmdmir}; the compact group sample is dominated by the optical red sequence, and the \mir canyon galaxies fall in a tight range on the optical red sequence, {\it not} in the green valley. These results are discussed in more detail below.

%%CMD%%%%%%%%%%%%%%%%%%%%%%%%%%%%%%%%%%%%%%%%%%%%%%
\subsection{Color-Magnitude Diagram}
The left panel in Figure \ref{cmdmir} shows the CMD for all {\ngal} compact group galaxies overlaid on contours representing the field sample from the VAGC. Comparison with the contours reveals that, relative to the field, compact group galaxies show a greater tendency to lie on the red sequence and form neither a distinct blue cloud nor green valley. As expected, the field galaxy sample forms a well-defined red sequence and blue cloud with the green valley wedged in between. The differences between the field and compact group samples are especially apparent in the histogram of $g-r$ color shown in Figure \ref{colorhist}. This figure illustrates that the two field samples are consistent, each form both a blue cloud and red sequence. Also shown in the histograms are the colors of galaxies from the Coma cluster \citep{mahajan10}. It is clear that the color distribution of compact group galaxies is similar to that of Coma galaxies, though the red sequence is stronger in Coma. Comparison with CMDs from \citet{hogg04} reveals that the structure of the compact group CMD is similar to that of very high density regions, which is not surprising as, like clusters, compact groups are also high density.

A relatively straightforward hypothesis that \mir transition galaxies in compact groups occupy the green valley in the optical CMD is clearly incorrect. Instead, the \mir transition galaxies occupy the red sequence rather than the green valley, spanning a tight optical color range $\left(0.73 < \left[g-r\right] < 0.77\right)$. The \mir active galaxies span the full optical color range, while \mir quiescent galaxies tend to fall along the optical red sequence with a slight excess at optically redder colors. Note that on the right in Figure \ref{cmdmir}, the \mir transition galaxies in LVL+SINGS span the range of optical colors.

%%Opt\mir%%%%%%%%%%%%%%%%%%%%%%%%%%%%%%%%%%%%%%%%%%%
\subsection{Optical vs \mir}
By comparing the optical and \mir color distributions of galaxies, we can determine how the properties that give rise to \mir transition galaxies relate to the evolution of star formation within galaxies. If \mir colors track star formation, we would expect them to lead optical colors. If \mir colors track heating of polycyclic aromatic hydrocarbons (PAHs), we expect them to follow optical colors.

The relationship between optical and \mir colors for compact group and LVL+SINGS galaxies, shown in Figure \ref{optmir}, has several notable features. Galaxies that are \mir quiescent are optically red. By comparison, \mir active galaxies span the full optical color range. As seen in Figure \ref{cmdmir}, compact group \mir transition galaxies fall on the optical red sequence. By comparison, the LVL+SINGS sample does contain optically blue \mir transition galaxies, which we see from Figure \ref{cmdmir} are relatively low magnitude (and likely low mass) galaxies. Their absence in compact groups is discussed in section \ref{evolveimply}. In both samples, we note a lack of galaxies that are blue in both \mir and optical colors; these galaxies likely recently experienced quenching, but would still have young stars present.

The total energy emitted at both optical and \mir wavelengths also provides insight into the global state of the galaxies. Using the SEDs of the galaxies, we determined the energy emitted at optical and infrared wavelengths by integrating $\nu L_\nu$ (over 4686\AA $<\lambda<$ 7480\AA and 3.6 $\mu$m $<\lambda<$ 8.0 $\mu$m). The resulting comparison for both the compact group sample and LVL+SINGS is shown in Figure~\ref{energy}. The \mir transition galaxies fall on a line indicating relative optical to \mir emission of 8.1 ($\sigma =1.6$) for the compact group sample and 7.5 ($\sigma =3.2$) for LVL+SINGS. Over the entire sample, the compact group galaxies show a larger scatter in this ratio than LVL+SINGS, especially for higher luminosities. The compact group sample shows a clear split in the energy plot between galaxies that are \mir active and \mir quiescent, with \mir quiescent galaxies tending to emit a larger fraction of energy at optical wavelengths. While not unexpected, this split is interesting because it is not nearly as apparent for LVL+SINGS where \mir quiescent galaxies are not clearly discernable from \mir transition or \mir active galaxies.

\input{cangals}

%%StarsvsDust%%%%%%%%%%%%%%%%%%%%%%%%%%%%%%%%%%%%%%
\subsection{Stars vs Dust}
The \mir canyon seen in compact groups also shows up clearly in a simple plot of 4.5 $\mu$m versus 8.0 $\mu$m luminosity (see Figure~\ref{luminosities}) as a dearth of galaxies on the diagonal line of unity. These particular \mir bands are of interest as the 8.0 $\mu$m band can be dominated by PAH emission, while the 4.5 $\mu$m band is typically dominated by stars and is virtually free of PAH emission. Thus, a direct comparison of these bands can serve as an indicator of the relative contribution of PAH emission to the SED of a galaxy. Of particular note is the fact that the \mir transition galaxies essentially lie on the unity line, indicating these galaxies are emitting equivalent energies in the 4.5 $\mu$m and 8.0 $\mu$m bands. For LVL+SINGS, the \mir transition galaxies also fall approximately on the line of unity. However, as shown previously in \citet{walker12} using other diagnostics, no dearth of galaxies in LVL+SINGS is apparent in the \mir canyon.

This comparison may indicate that the rapid transition of compact group galaxies through the \mir canyon is strongly related to their PAH emission. This finding is consistent with the hypothesis that the \mir canyon in the compact group sample is due to global star formation being truncated on very short timescales. Alternately, the canyon being tied to PAH emission could indicate modest accretion events from the intragroup medium that provide small amounts of gas and dust for a short period of time.

%%%%%%%%%%%%%%%%%%%%%%%%%%%%%%%%%%%%%%%%%%%%%%%%%%%
%	4. Discussion
%%%%%%%%%%%%%%%%%%%%%%%%%%%%%%%%%%%%%%%%%%%%%%%%%%%
\section{DISCUSSION}
%%Transition Galaxies%%%%%%%%%%%%%%%%%%%%%%%%%%%%%%%%%%%%%
\subsection{Transition Galaxies in Optical and \mir}\label{optmirdiscussion}
As the green valley is considered to be a ``transition region'' between star forming galaxies in the blue cloud and passively evolving galaxies on the red sequence, a straightforward hypothesis would be that this region of the optical CMD corresponds to the \mir canyon. However, this hypothesis is clearly incorrect, as the tight range of optical colors occupied by the \mir transition galaxies in compact groups corresponds to the optical red sequence. Thus, \mir transition galaxies in compact groups have optical colors that indicate a ``red and dead'' population, a population with significant reddening from dust, or perhaps some combination thereof. However, if dust were causing significant reddening in a star-forming galaxy, then the galaxy would be very red in the \mir. Visual inspection indicates a smooth optical morphology with no indication of dust lanes or irregularities.

One possible interpretation is that the star formation in \mir transition galaxies has recently turned off (given the red optical colors), but there is still sufficient stellar emission to generate dust and PAH emission. The optical colors of \mir transition galaxies in compact groups imply that the optical transition precedes the \mir transition. This relative sequence suggests that the dust and PAHs are not being heated by the most massive stars. This is further supported by the fact that SEDs show the dust in \mir transition galaxies to be at an intermediate temperature \citep[see Figure 13 of][]{walker12}; if these are galaxies that no longer have massive stars to heat the dust, we would expect to see the dust cooling and \mir emission declining.

%%Implications for Galaxy Evolution%%%%%%%%%%%%%%%%%%%%%%%%%%%%%%%
\subsection{Implications for Galaxy Evolution}\label{evolveimply}
Of particular note is the fact that for the sample of compact group galaxies, the \mir canyon {\it does not} correspond to the optical green valley, which indicates that \mir transition galaxies have already transitioned to the optical red sequence. There are clear indications in the compact group CMD that galaxy evolution in compact groups is markedly different than secular evolution in the field. The increased fraction of galaxies with redder optical colors (a similarity with the Coma CMD), combined with the fact that all of the \mir quiescent galaxies fall in the optical red sequence, seems to indicate that compact group galaxies on the whole are more evolved than field galaxies. Thus, either the compact group environment is more like the cluster environment than the field environment, or galaxies in clusters experience pre-processing in a compact group-like environment \citep{cortese06}.

Galaxies in clusters are known to have older stellar populations on average than those in the field \citep[e.g.][]{bernardi98,sanchez06,gobat08}. Optical studies of cluster galaxies show strong red sequences; in fact the optical CMD for the Coma cluster \citep{mahajan10} is remarkably similar to our compact group CMD. Thus, galaxy processing appears to be important even in these smaller structures. This is supported by recent findings that even slight increases in local density or small groups lead to older populations \citep{blanton07a,blanton09}. Compact groups, though small, show densities comparable to the centers of clusters \citep{hickson82}.

\begin{figure}
\plotone{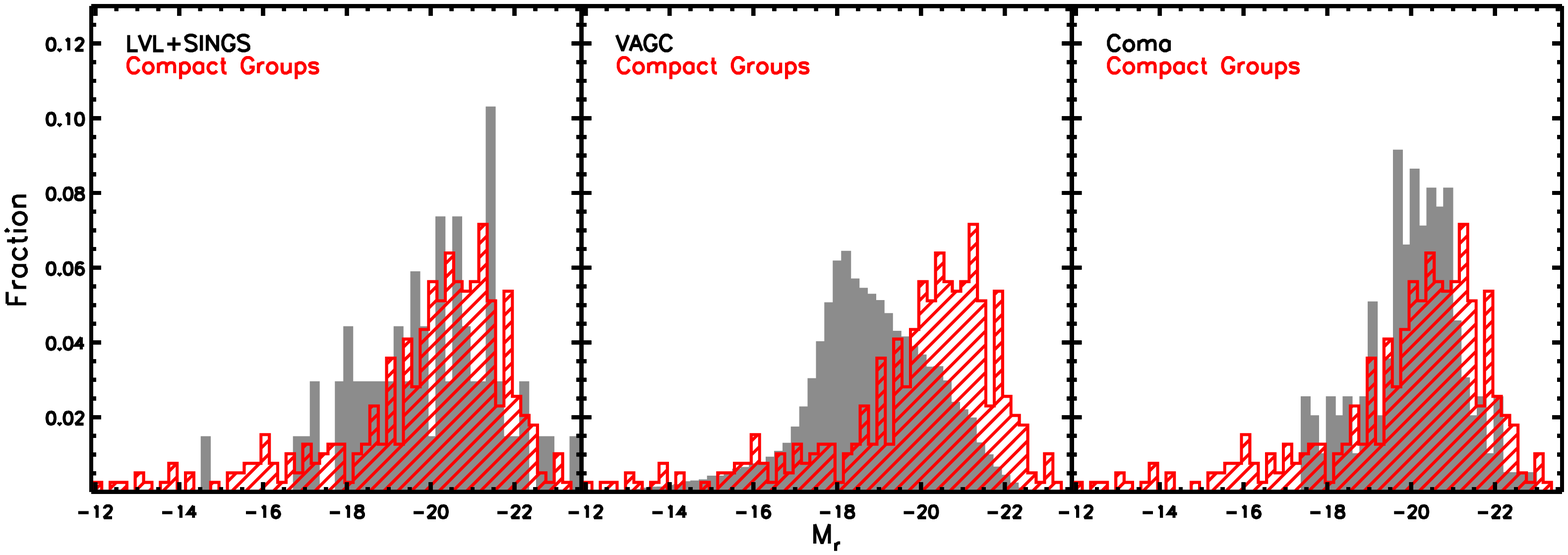}
\caption{Comparison of the absolute magnitudes of the compact group sample and our comparison samples. The LVL+SINGS and compact group samples are well matched, both including more low-luminosity galaxies than Coma. The difference with VAGC is pronounced, and comparison at $M_r < -19$ should be done with care.}
\label{lumhist}
\end{figure}

\begin{figure}
\plottwo{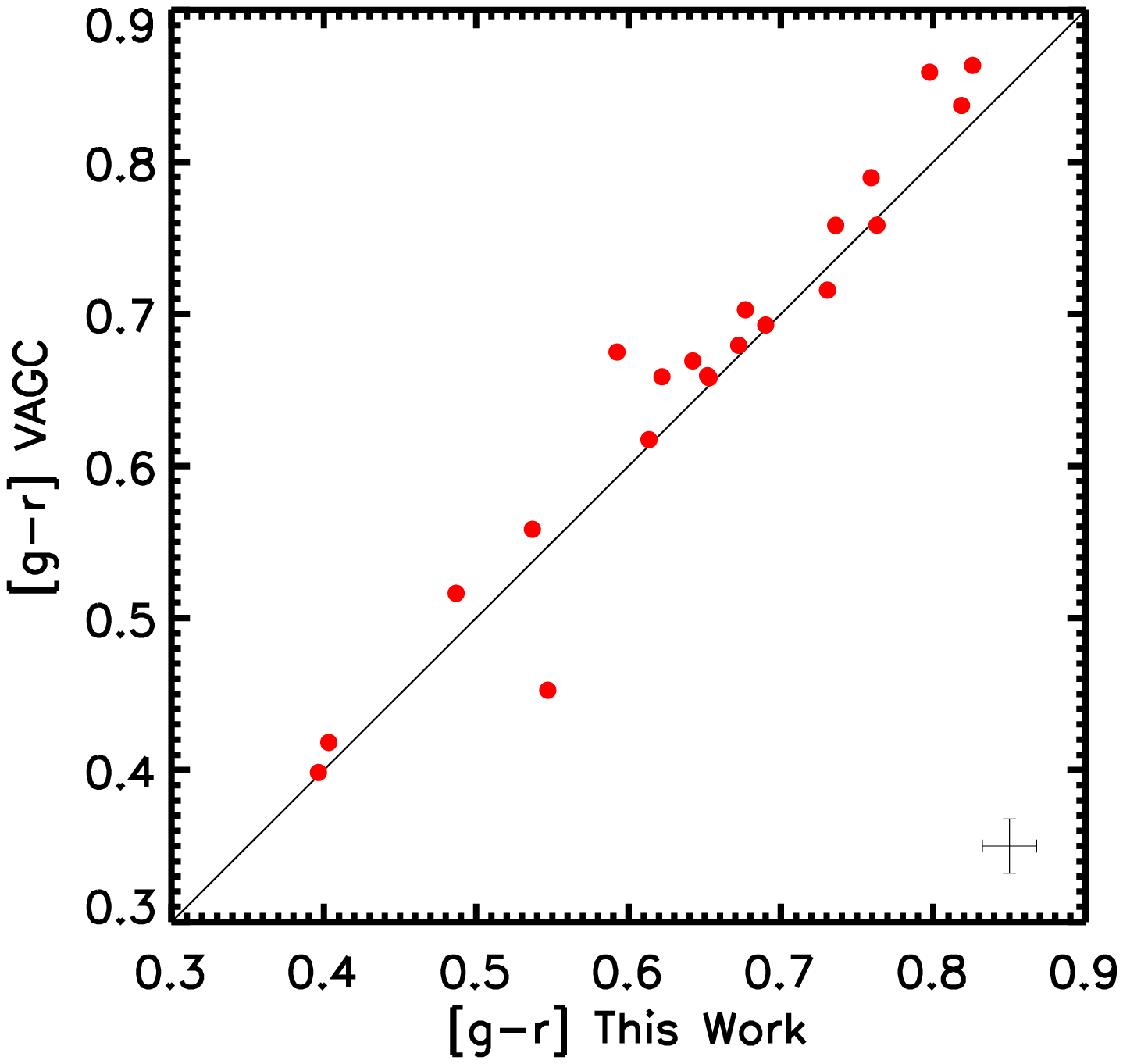}{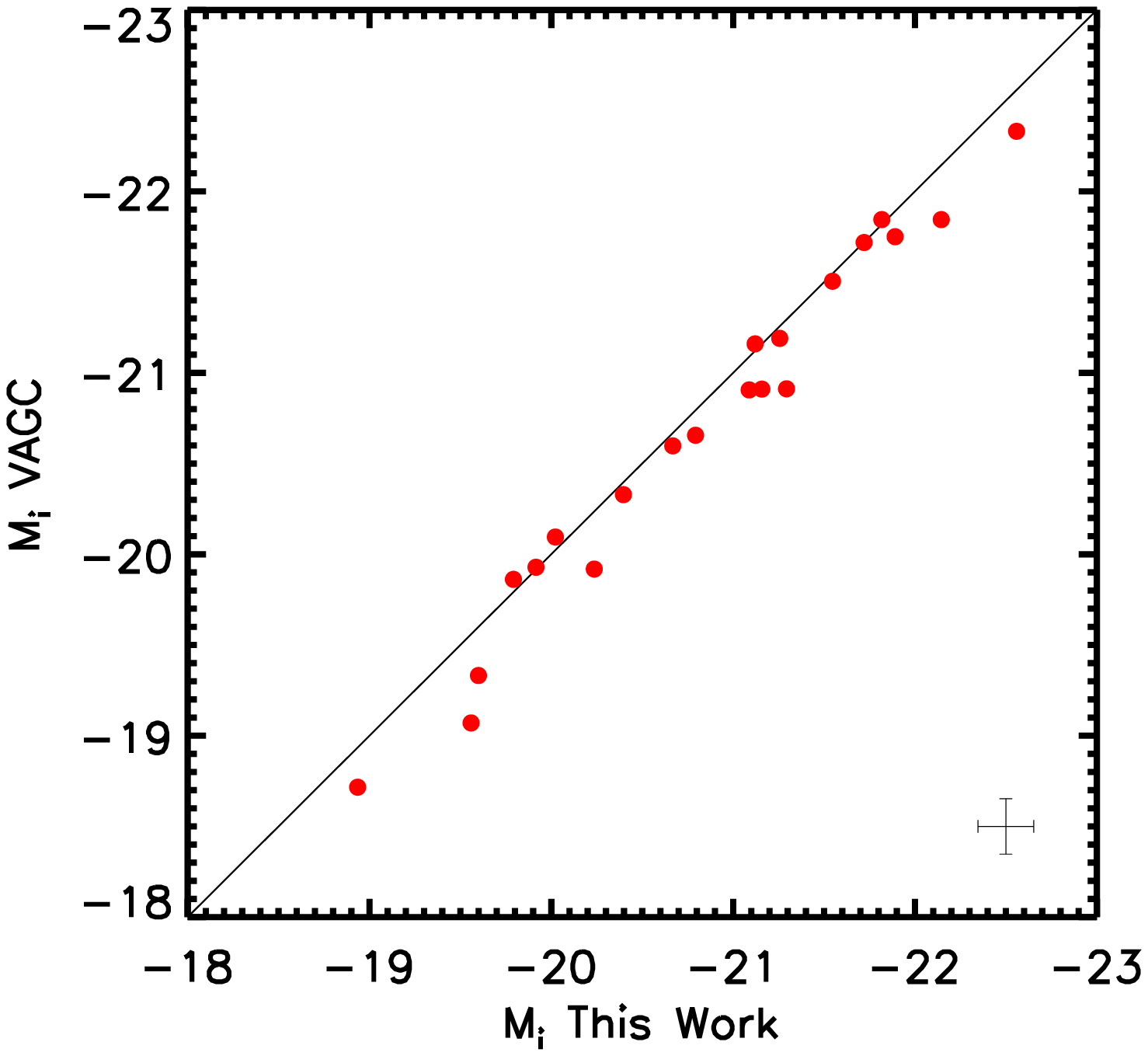}
\caption{Comparison of our photometry with the photometry from the low-z NYU VAGC showing that the two methods are consistent. The error bars indicate the average error for our data, the line indicates equality.}
\label{vssloan}
\end{figure}

\begin{figure}
\plottwo{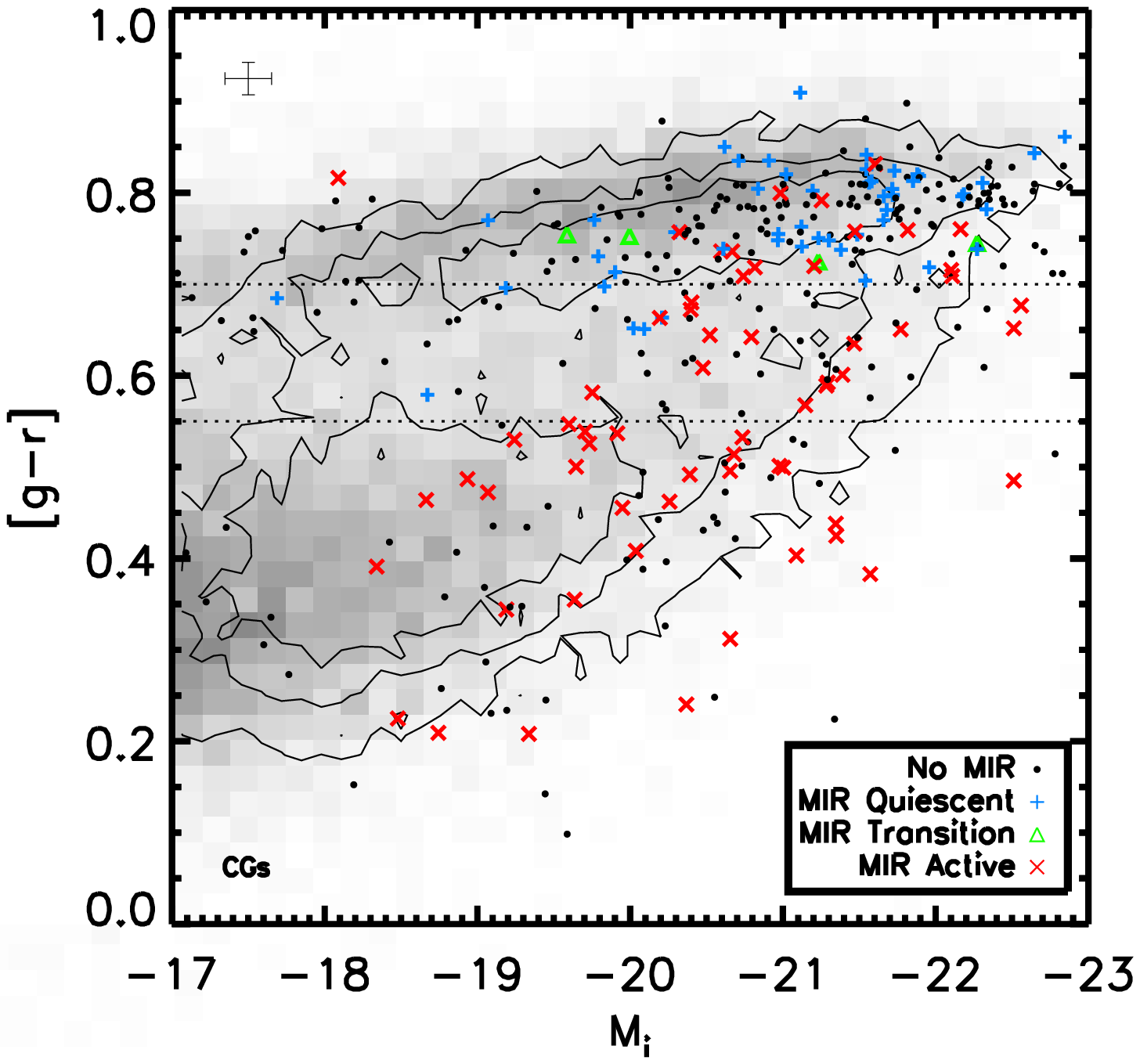}{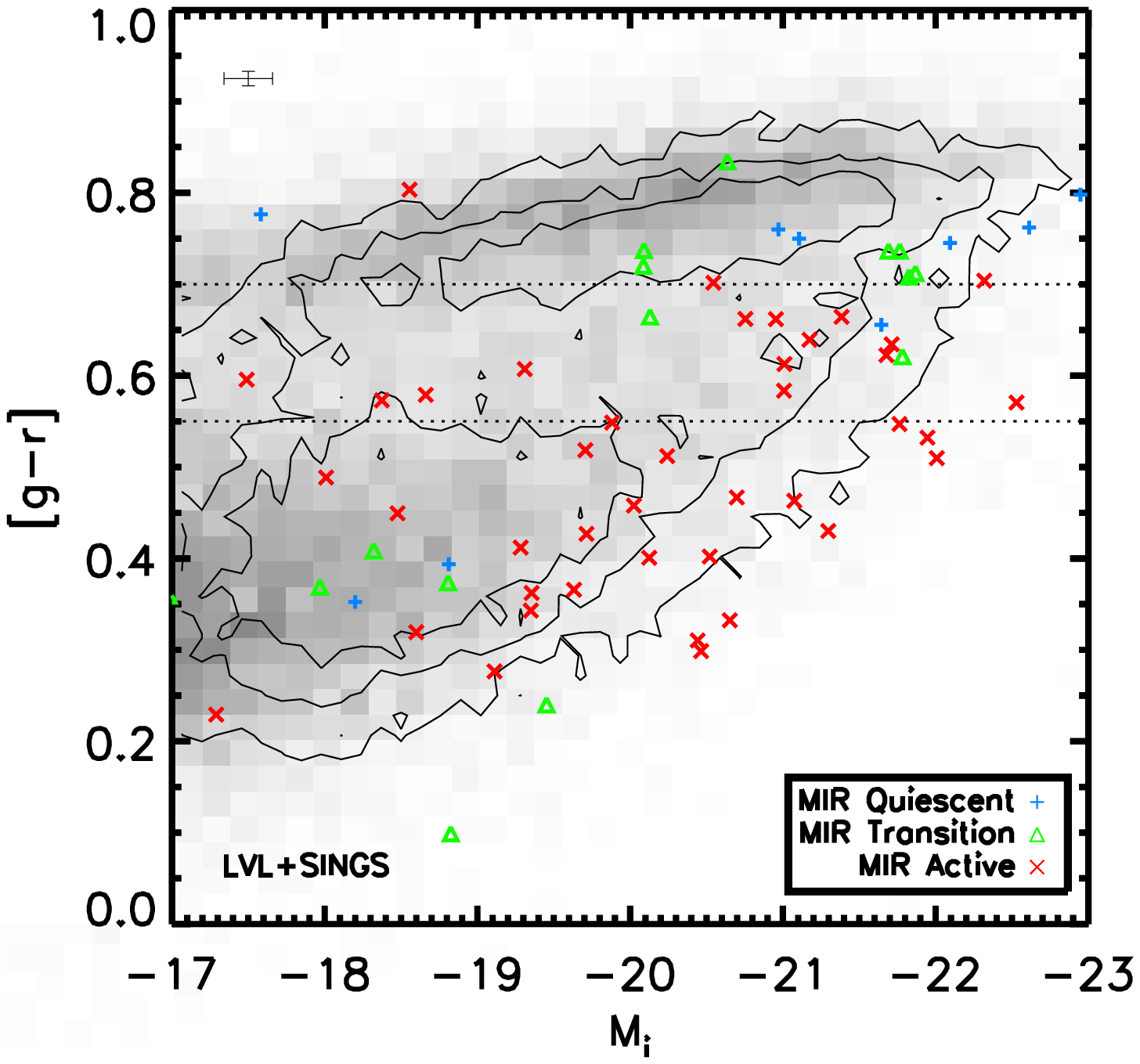}
\caption{Color-magnitude diagram of the {\it left}: compact group sample and {\it right}: LVL+SINGS sample \citet{dale09} (individual points) overlaid on the VAGC \citet{blanton05b} (contours and grayscale). The symbols are colored by which region of \mir color space they fall in from \citet{walker12}. The dotted lines indicate the bounds of the green valley. We see that the compact group sample is dominated by the optical red sequence. In compact groups, the \mir canyon galaxies fall in a tight range on the optical red sequence while they span the range of optical colors for LVL+SINGS.}
\label{cmdmir}
\end{figure}

\begin{figure}
\plottwo{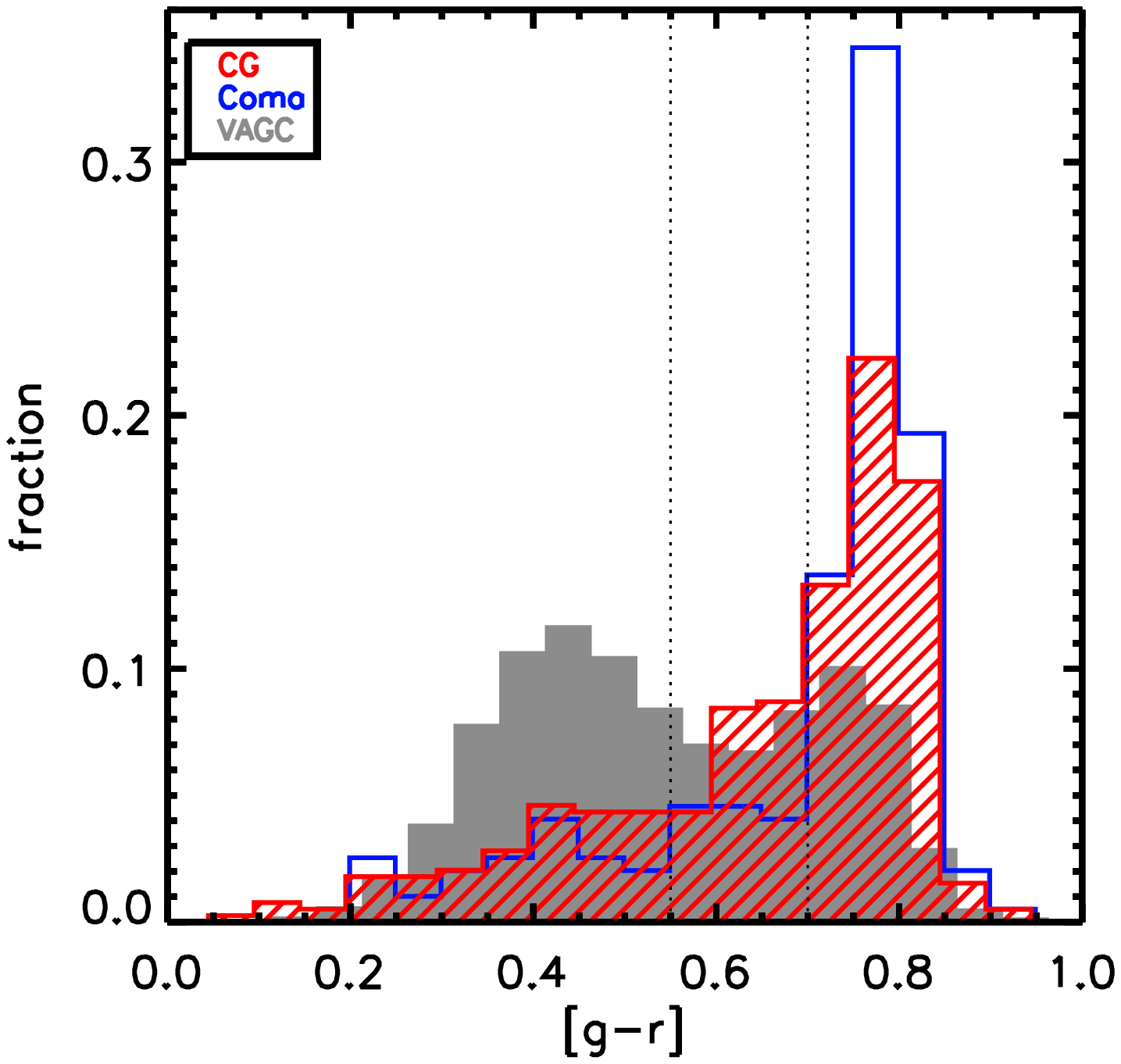}{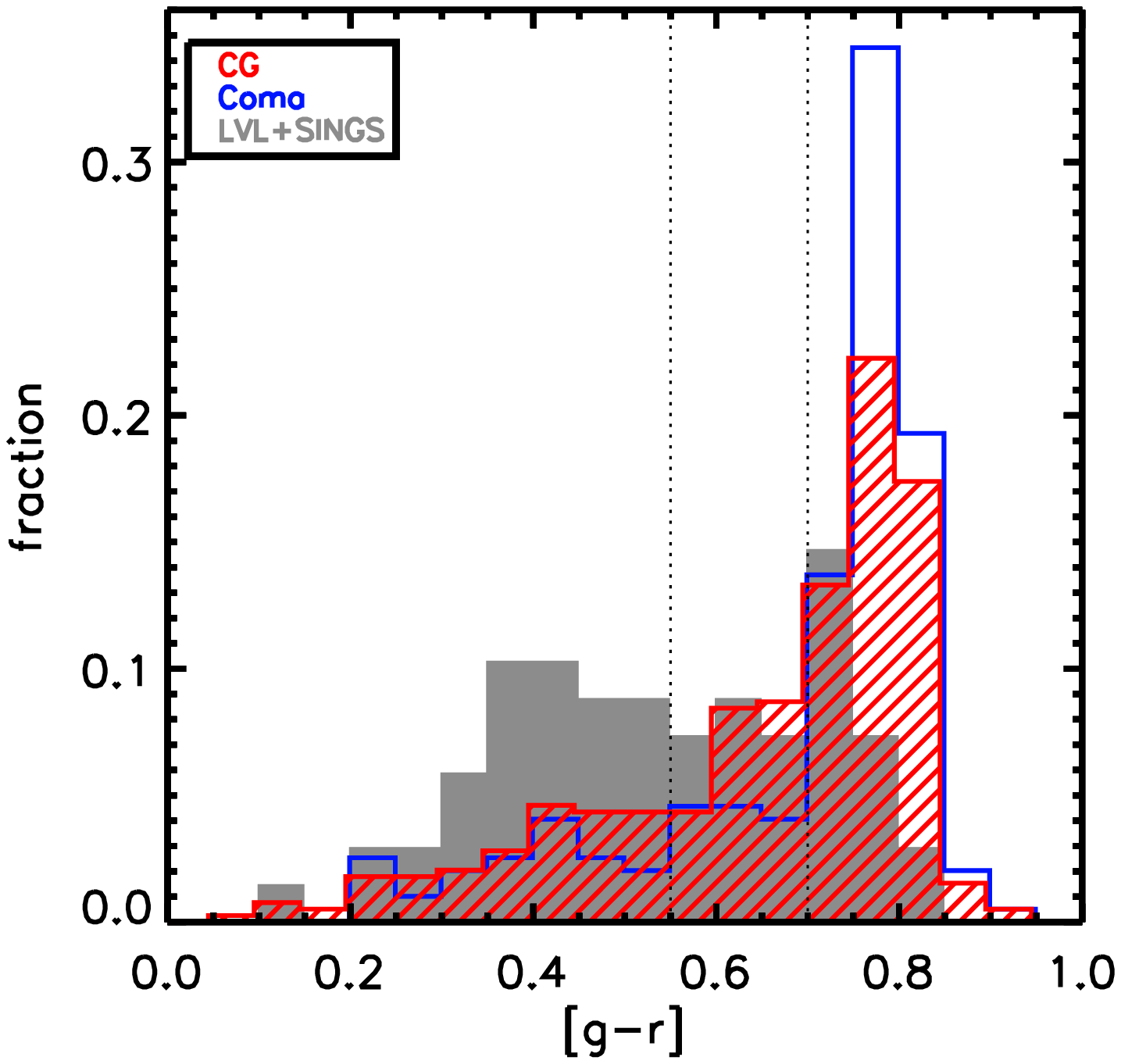}
\caption{Histogram of g-r colors for the compact group (red stripes) and Coma samples from \citet{mahajan10} (blue line) overlaid on the {\it left}: VAGC \citet{blanton05b} and {\it right}: LVL+SINGS samples(grey solid). This clearly illustrates the dominance of the red sequence in compact groups and that our two field samples are consistent.}
\label{colorhist}
\end{figure}

\begin{figure}
\plottwo{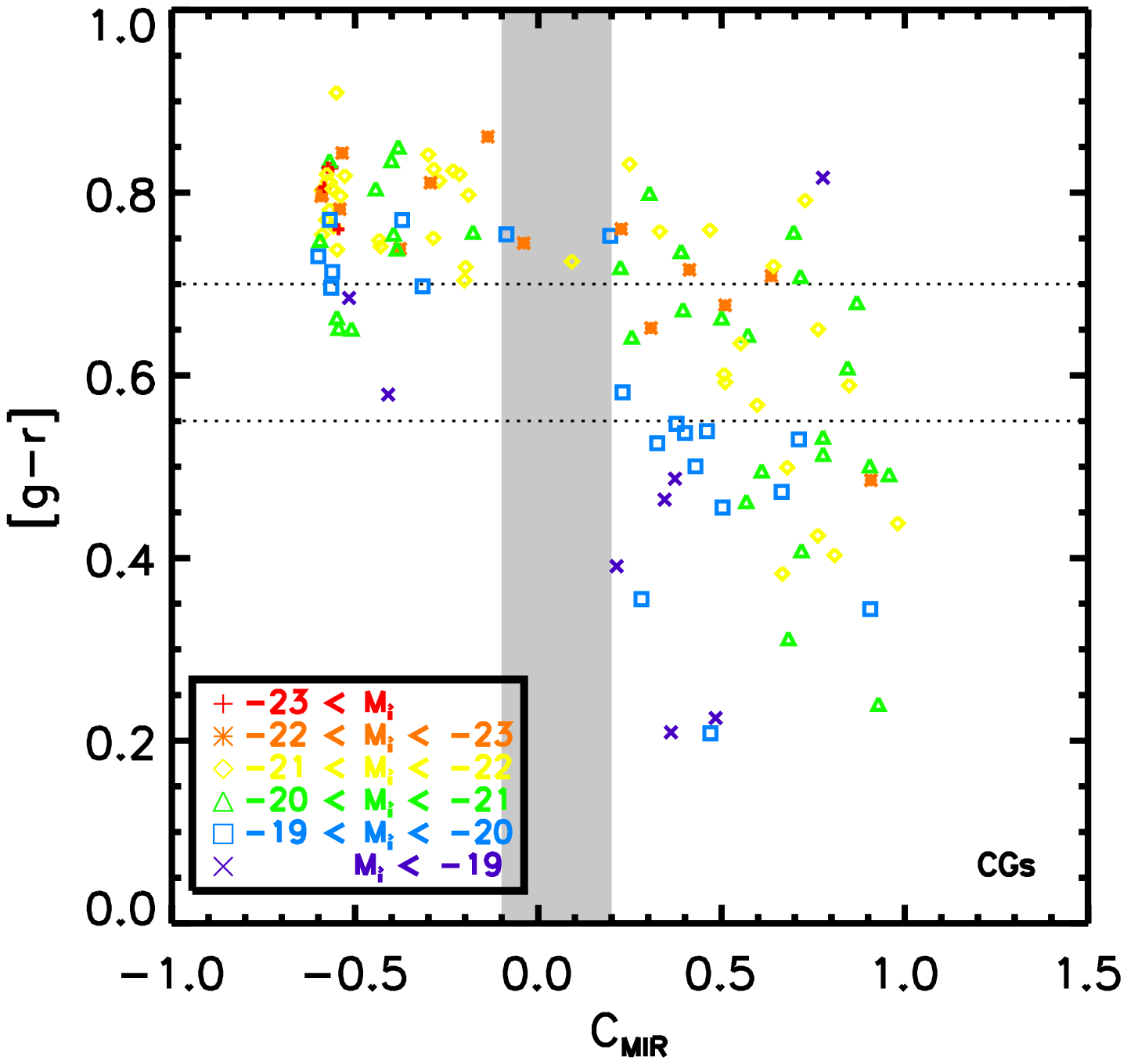}{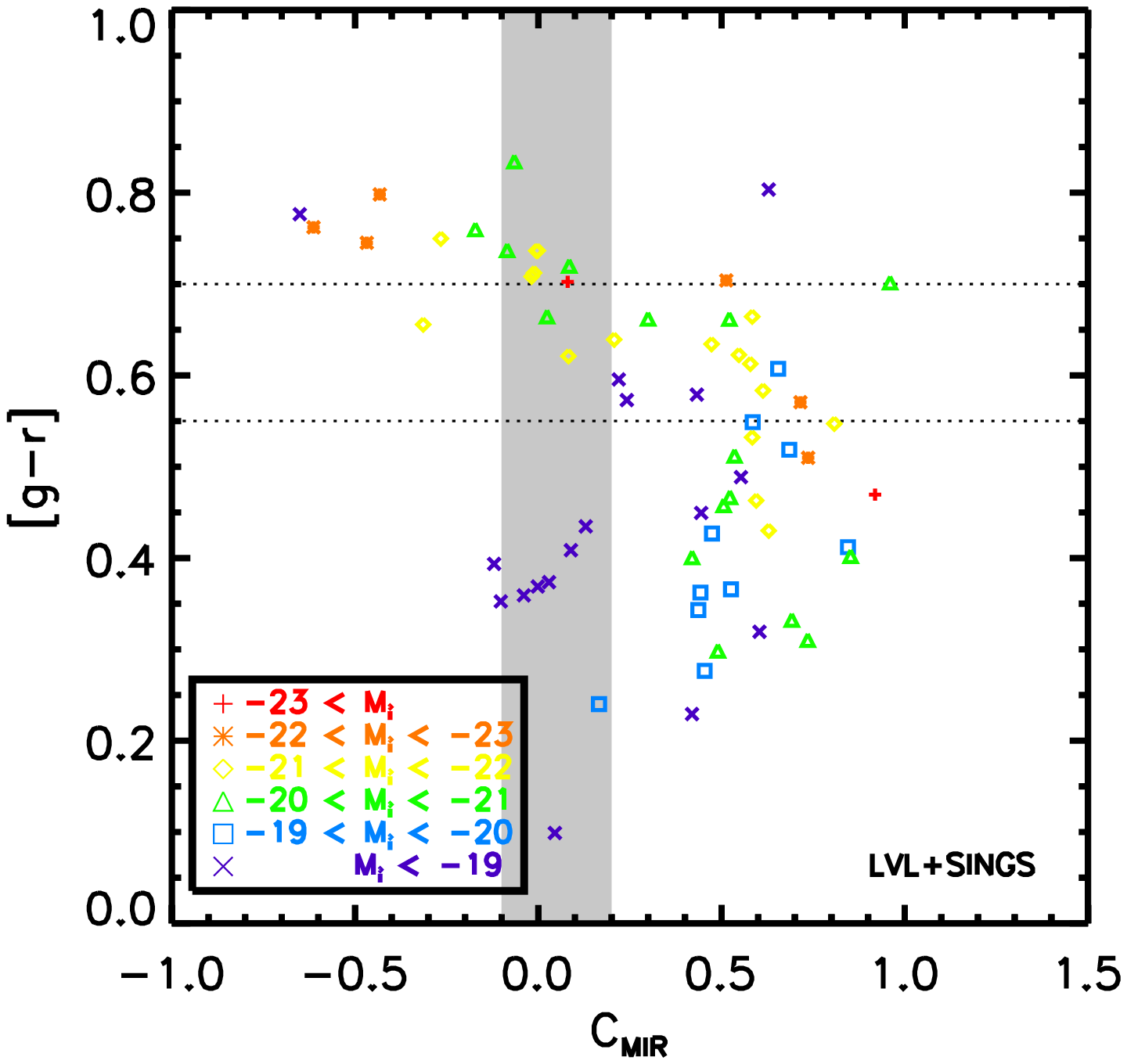}
\caption{ Comparison of optical with \mir colors with symbols indicating {\it i}-band magnitude, $\mathrm{M_i}$ for {\it left}: compact groups and {\it right}: LVL+SINGS. $\mathrm{C_{MIR}}$ indicates \mir color along the curve from Figure 3 of \citet{walker12}. The gray stripe indicates the canyon region from \citet{walker12}; the green valley is between the dotted lines. The \mir quiescent galaxies that are optically red, while \mir active galaxies span the entire range of optical colors. The \mir transition galaxies all fall on the red sequence in compact groups, but in LVL+SINGS they span the full range of optical colors.}
\label{optmir}
\end{figure}

\begin{figure}
\plottwo{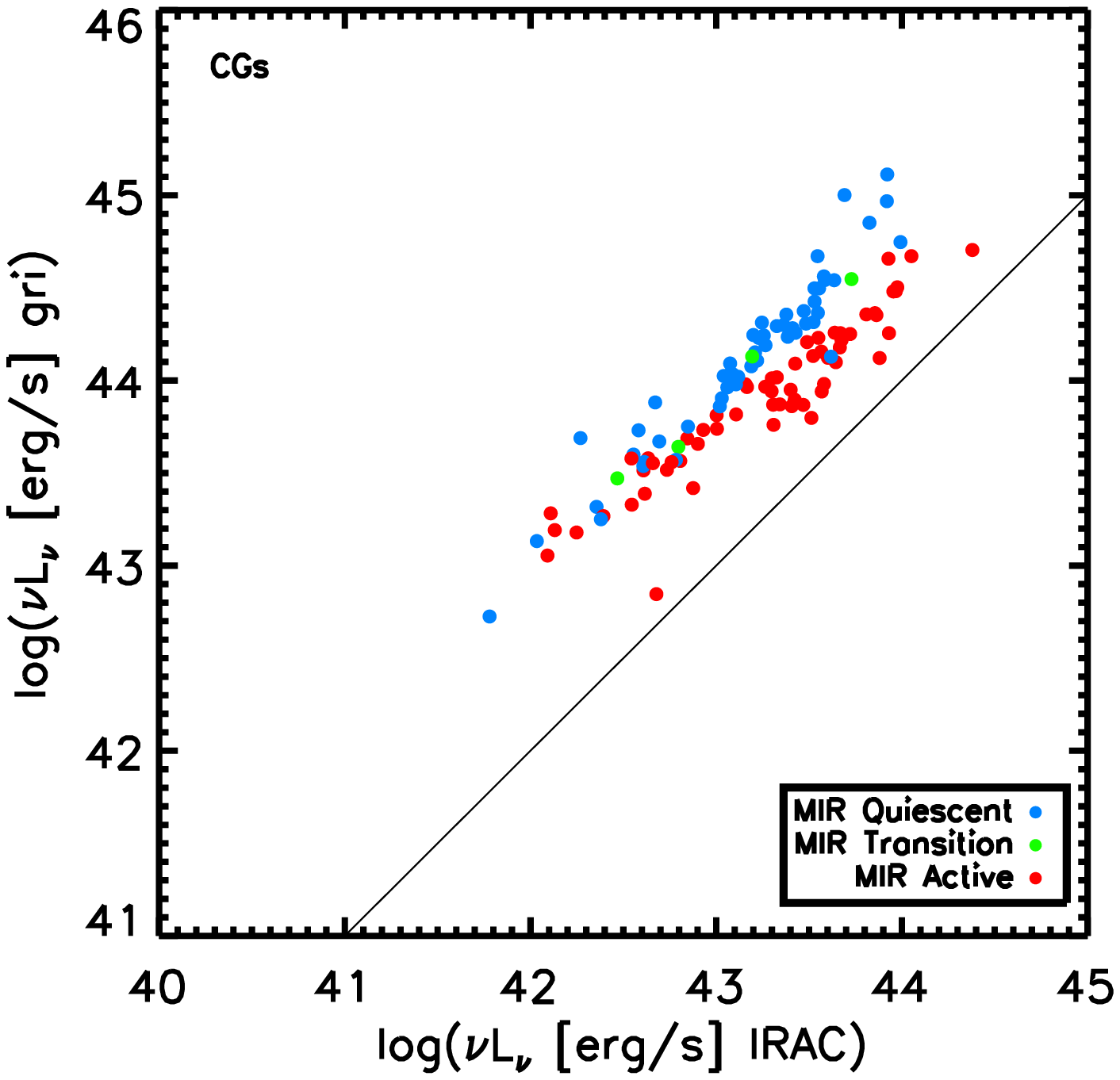}{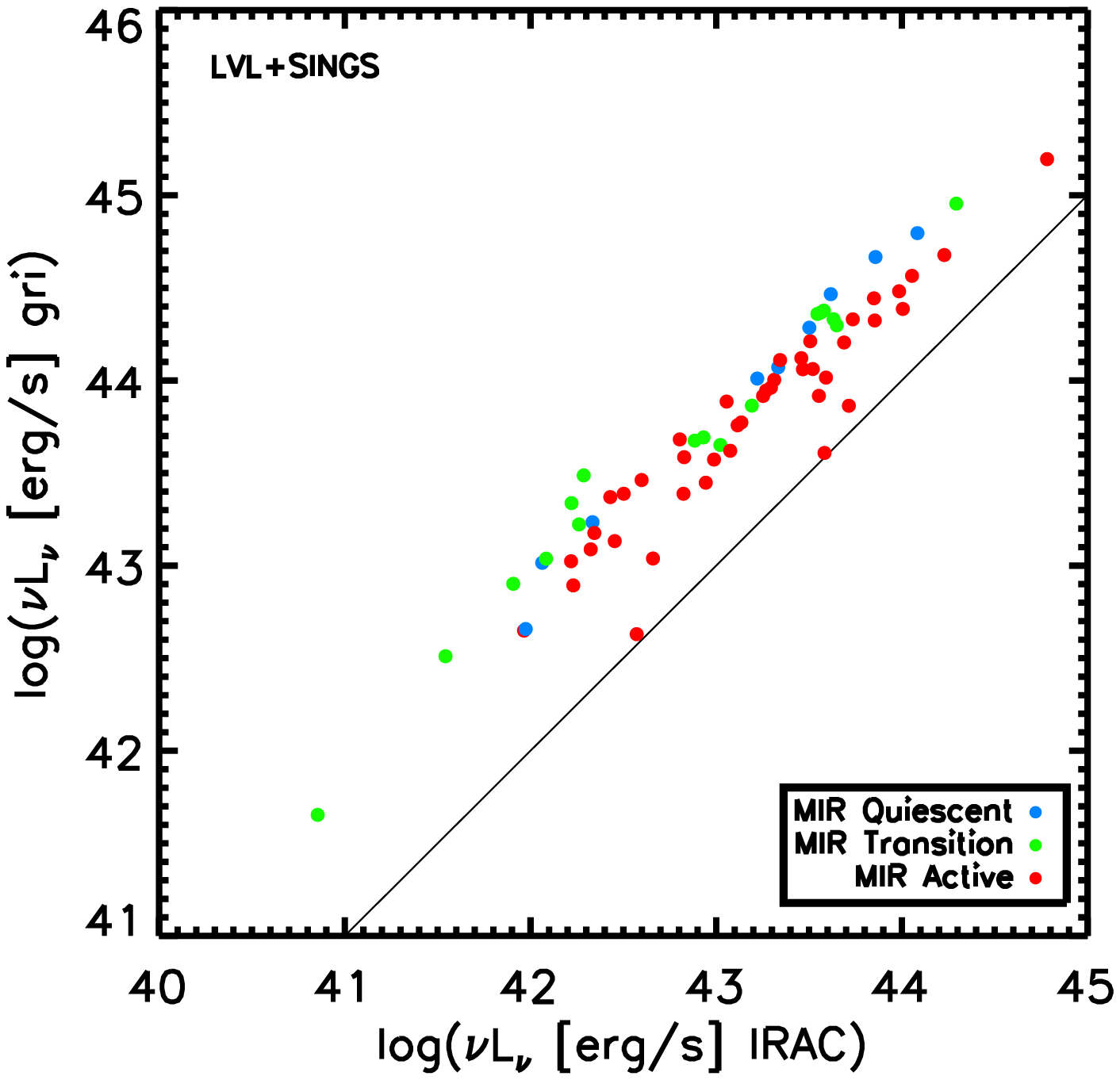}
\caption{A comparison of the amount of light emitted in the optical vs the \mir for {\it left}: compact group galaxies and {\it right}: LVL+SINGS galaxies. The line indicates equality. We see a clear split between \mir active and \mir quiescent galaxies for compact groups at $\mathrm{\nu L_\nu} >\sim10^{43}\;\unit{erg/s}$ with an offset greater than the typical errors of $\sim10\%$, with \mir quiescent galaxies tending to emit a larger fraction of energy at optical wavelengths.}
\label{energy}
\end{figure}

\begin{figure}
\plottwo{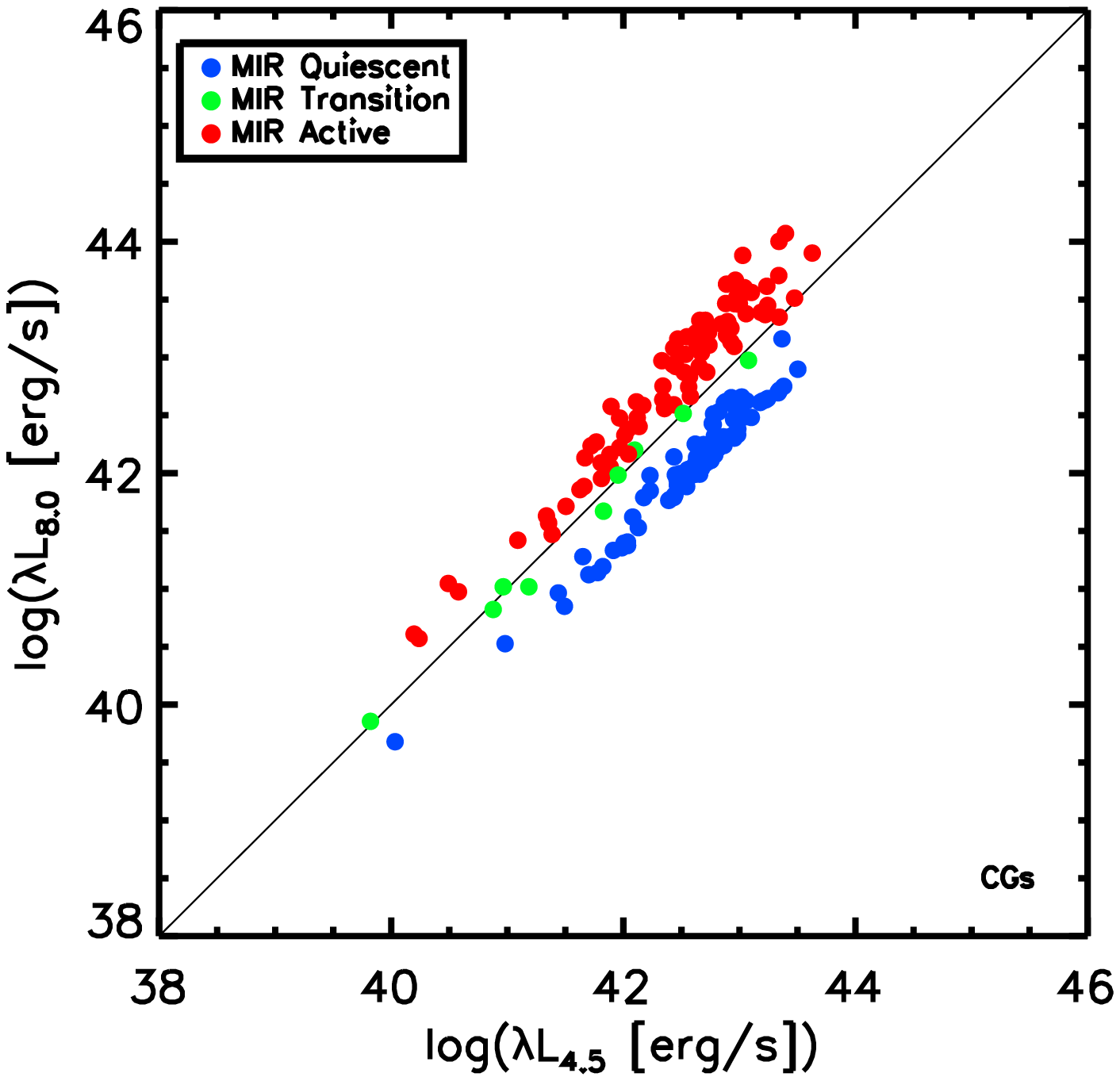}{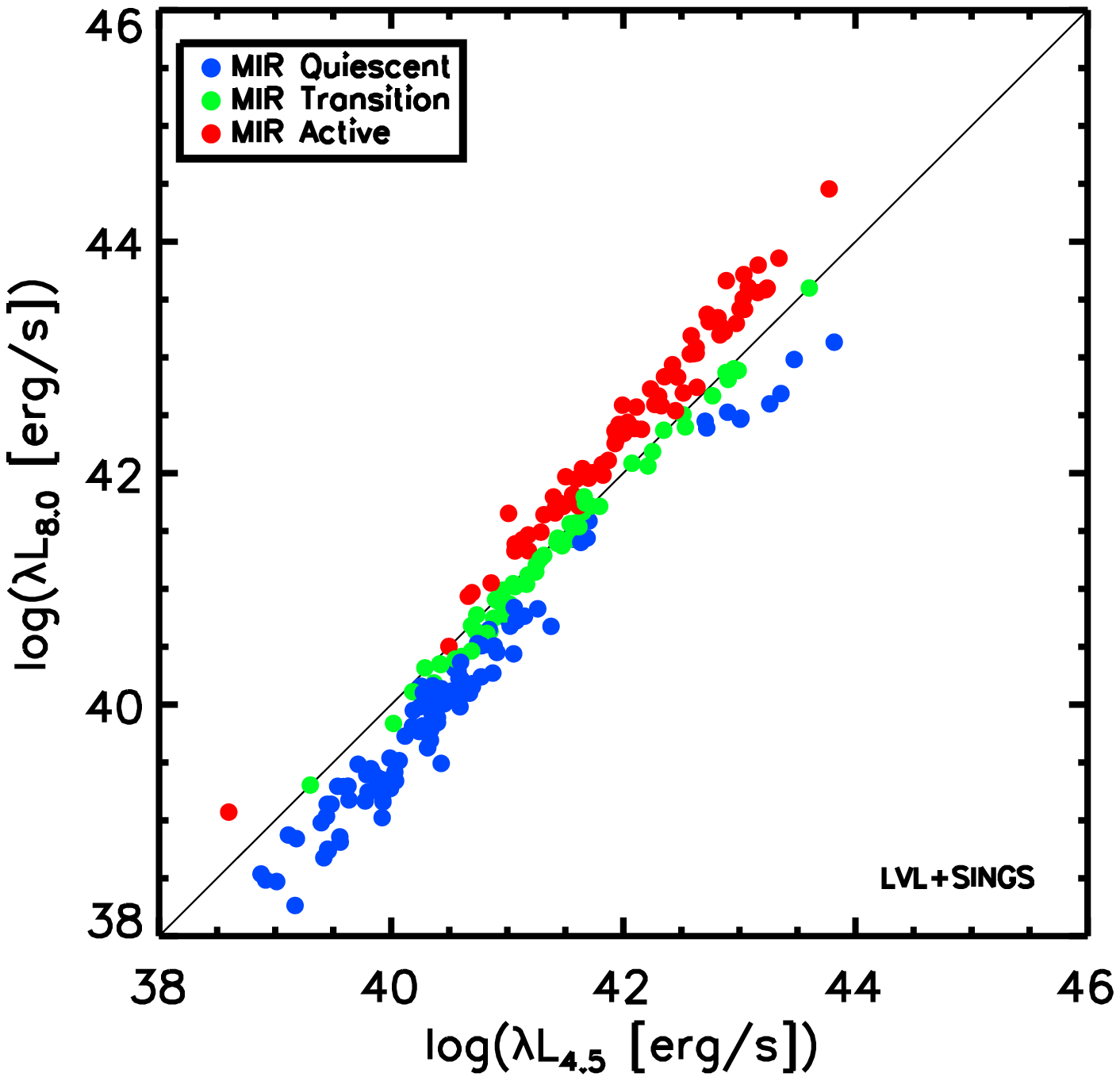}
\caption{A comparison of the 8.0 $\mu$m and 4.5 $\mu$m luminosities for {\it left}: compact group galaxies and {\it right}: LVL+SINGS galaxies. Note that \mir transition galaxies fall on the unity line, indicating these galaxies are emitting equivalent energies in the 4.5 $\mu$m (dominated by stars) and 8.0 $\mu$m (potentially dominated by PAHs) bands. We also note the separation of the \mir-quiescent and \mir-active galaxies in this plot, indicating that the \mir bimodality is not due to behavior manifest in a single band.}
\label{luminosities}
\end{figure}

A comparison of the CMDs for the compact group and LVL+SINGS galaxies  leads to two scenarios. The first is that the processes involved in the decline of dust emission take place on faster timescales in compact groups than in field galaxies, but this is complicated by the presence of the low-mass, optically blue \mir transition galaxies in the LVL+SINGS sample. Furthermore, given how well matched the compact group and LVL+SINGS samples are in $\mathrm{M_r}$ as shown in Figure \ref{lumhist}, we find it unlikely that the lack of \mir transition galaxies in compact groups could be due to completeness issues. The second is that the dearth of galaxies in the \mir canyon {\it may} be attributed to a lack of moderately star forming, low mass galaxies. This hypothesis could be explained by compact groups being inhospitable to these galaxies; they either lose their gas very easily or are accreted by larger members of the group. The lack of low mass star-forming galaxies could in part also be due to selection -- there is a lower limit to galaxies identified as belonging to the group, though we do not see \mir canyon in LVL+SINGS, which has a similar population of galaxies, as shown in Figure \ref{lumhist}. Lower mass, star-forming galaxies are likely to be farther away from the core, and so again could be missed in the selection, like the star-forming dwarfs found on the outskirts of HCG 59 \citep{konstantopoulos12}. It is not clear which effect dominates, the dearth of \mir transition galaxies is likely due to some combination of these factors.

The compact group environment appears to impact galaxy evolution by being inhospitable to moderate levels of star formation; the environment either enhances or terminates star formation within member galaxies. It may also be that star formation is first enhanced in member galaxies, then terminated when the star forming gas is removed due to group interactions. Therefore, how gas is processed (e.g. through star formation or stripping) affects the properties of compact group galaxies. Thus these results indicate that the location and phase of gas in compact groups plays a crucial role in determining the evolution of compact group galaxies.

%%%%%%%%%%%%%%%%%%%%%%%%%%%%%%%%%%%%%%%%%%%%%%%%%%%
%	Appendix - Photometry
%%%%%%%%%%%%%%%%%%%%%%%%%%%%%%%%%%%%%%%%%%%%%%%%%%%
\appendix
\section{Photometry}
Tables \ref{hcggals} and \ref{rscggals} present the apparent magnitudes of the sample, along with alternate names for the RSCG galaxies.
\input{hcggals}
\input{rscggals}

%%%%%%%%%%%%%%%%%%%%%%%%%%%%%%%%%%%%%%%%%%%%%%%%%%%
%	Figures
%%%%%%%%%%%%%%%%%%%%%%%%%%%%%%%%%%%%%%%%%%%%%%%%%%%

%%%%%%%%%%%%%%%%%%%%%%%%%%%%%%%%%%%%%%%%%%%%%%%%%%%
%	Acknowledgements
%%%%%%%%%%%%%%%%%%%%%%%%%%%%%%%%%%%%%%%%%%%%%%%%%%%
\acknowledgments
Funding for the Sloan Digital Sky Survey (SDSS) has been provided by the Alfred P. Sloan Foundation, the Participating Institutions, the National Aeronautics and Space Administration, the National Science Foundation, the U.S. Department of Energy, the Japanese Monbukagakusho, and the Max Planck Society. The SDSS Web site is http://www.sdss.org/.

The SDSS is managed by the Astrophysical Research Consortium (ARC) for the Participating Institutions. The Participating Institutions are The University of Chicago, Fermilab, the Institute for Advanced Study, the Japan Participation Group, The Johns Hopkins University, Los Alamos National Laboratory, the Max-Planck-Institute for Astronomy (MPIA), the Max-Planck-Institute for Astrophysics (MPA), New Mexico State University, University of Pittsburgh, Princeton University, the United States Naval Observatory, and the University of Washington.

This research has made use of the NASA/IPAC Extragalactic Database (NED) which is operated by the Jet Propulsion Laboratory, California Institute of Technology, under contract with the National Aeronautics and Space Administration.

We thank the referee for their constructive comments to improve the paper.

{\it Facilities:} \facility{Sloan}

%%%%%%%%%%%%%%%%%%%%%%%%%%%%%%%%%%%%%%%%%%%%%%%%%%%
%	Bibliography
%%%%%%%%%%%%%%%%%%%%%%%%%%%%%%%%%%%%%%%%%%%%%%%%%%%

\end{document}